\documentclass[a4paper,twoside]{article}

\usepackage{amsmath,amsthm, amsfonts, amssymb, amsxtra,amsopn}
\usepackage{pgfplots}
\pgfplotsset{compat=1.13}
\usepackage{pgfplotstable}
\usepackage{graphicx}
\usepackage{multirow}
\usepackage{tabu}
\usepackage{algorithmic}
\usepackage{longtable}
\usepackage{booktabs}
\usepackage{listings}
\usepackage{cmap}
\usepackage{colortbl}
\usepackage{adjustbox}
\usepackage{epsfig}

\usepackage{subfigure}

\PassOptionsToPackage{hyphens}{url}

\usepackage{hyperref}
\hypersetup{colorlinks=true,linkcolor=black,citecolor=black,urlcolor=blue,filecolor=black}
\hypersetup{pdfpagemode=UseNone,pdfstartview=}

\usepackage[tableposition=top]{caption}
\usepackage{enumitem}
\setlist[itemize]{noitemsep, topsep=0pt}

\usepackage{SCITEPRESS}     

\def\z{\phantom{0}}

\def\fs{\footnotesize}
\def\un{\underline{\phantom{M}}}

\def\naive{na\"{i}ve}

\pgfkeys{
    /pgf/number format/fixed zerofill=true }
    
\pgfplotstableset{
    color cells/.code={%
        \pgfqkeys{/color cells}{#1}%
        \pgfkeysalso{%
            postproc cell content/.code={%
                \begingroup
                \pgfkeysgetvalue{/pgfplots/table/@preprocessed cell content}\value
\ifx\value\empty
\endgroup
\else
                \pgfmathfloatparsenumber{\value}%
                \pgfmathfloattofixed{\pgfmathresult}%
                \let\value=\pgfmathresult
                \pgfplotscolormapaccess
                    [\pgfkeysvalueof{/color cells/min}:\pgfkeysvalueof{/color cells/max}]%
                    {\value}%
                    {\pgfkeysvalueof{/pgfplots/colormap name}}%
                \pgfkeysgetvalue{/pgfplots/table/@cell content}\typesetvalue
                \pgfkeysgetvalue{/color cells/textcolor}\textcolorvalue
                \toks0=\expandafter{\typesetvalue}%
                \xdef\temp{%
                    \noexpand\pgfkeysalso{%
                        @cell content={%
                            \noexpand\cellcolor[rgb]{\pgfmathresult}%
                            \noexpand\definecolor{mapped color}{rgb}{\pgfmathresult}%
                            \ifx\textcolorvalue\empty
                            \else
                                \noexpand\color{\textcolorvalue}%
                            \fi
                            \the\toks0 %
                        }%
                    }%
                }%
                \endgroup
                \temp
\fi
            }%
        }%
    }
}

\usepackage{multicol}
\usepackage{pslatex}
\usepackage{apalike}

\usepackage{SCITEPRESS}

\begin{document}

\title{A Comparative Analysis of Android Malware}

\author{\authorname{Neeraj Chavan\sup{1}, Fabio Di Troia\sup{1} and Mark Stamp\sup{1}}
\affiliation{\sup{1}Department of Computer Science, San Jose State University, San Jose, California, USA}
\email{neerajpadmakar.chavan@sjsu.edu, fabioditroia@msn.com, mark.stamp@sjsu.edu}
}


\abstract{In this paper, we present a comparative analysis of benign and
malicious Android applications, based on static features. In particular, we focus
our attention on the permissions requested by an application. 
We consider both binary classification of malware versus benign, as well as
the multiclass problem, where we classify malware samples into their
respective families. Our experiments are based on substantial malware datasets 
and we employ a wide variety of machine learning techniques, including
decision trees and random forests, support vector machines, logistic model trees, 
AdaBoost, and artificial neural networks. We find that permissions are a strong
feature and that by careful feature engineering, we can significantly reduce the 
number of features needed for highly accurate detection and classification.}

\onecolumn \maketitle \normalsize \setcounter{footnote}{0} \vfill

\section{\uppercase{Introduction}}\label{sect:intro}

\noindent As of~2017, the Android OS accounted for more than~85\%\ of the mobile OS market,
and there were more than~82 billion application (app) downloads from the Google Play
store during~2017~\cite{AndroidTop10}.
Predictably, the number of Android malware apps is also large---it 
is estimated that there were~3.5 million such apps in~2017, 
representing more than a six-fold increase since~2015~\cite{naked}. 
It follows that effective malware detection on Android 
devices is of critical importance.

Features that can be collected without executing the code
are said to be ``static,'' while features that require execution (or emulation)
are considered ``dynamic.''  Dynamic analysis is generally more informative
and dynamic features are typically more difficult for malware writers
to defeat via obfuscation~\cite{DamodaranTVAS17}. 
However, extracting dynamic features is likely
to be far more costly and time consuming, as compared to most static features.
Since efficiency is important on a mobile platform, in this paper,
we focus on static analysis. Specifically, we consider the related 
problems of Android malware detection and classification based on 
requested permissions---features that are easily obtained from the manifest file.
We analyze this feature over substantial malware datasets, and we consider the problem 
of feature reduction in some detail. Perhaps somewhat surprisingly, 
we find that a small subset of these features suffices. This is
significant, since malware detection on a mobile device must be fast, efficient,
and effective---the approach consider here meets all three of these criteria.

The remainder of this paper is organized as follows. In Section~\ref{sect:back},
we briefly discuss relevant background topics, including related work.
In Section~\ref{sect:exp} we discuss our experimental design and give our
experimental results. We also provide some discussion of our results.
Finally, in Section~\ref{sect:con} we conclude the paper and give some suggestions
for future work.

\section{\uppercase{Background}}\label{sect:back}

\noindent In this section, we first briefly discuss relevant background topics.
First, we outline each of the machine learning techniques
considered in this paper. Then we discuss some examples of relevant
related work.

\subsection{Machine Learning Techniques}\label{sect:ML_techs}

In this research, we employ a wide variety of machine learning techniques.
A detailed discussion of these techniques is well beyond the scope of this
paper---here, we simply provide a high-level overview.

\begin{description}
\item[Random forest] can be viewed as a generalization of the simple concept
of a decision tree~\cite{BreimanCutler2001}. 
While decision trees are indeed simple, they tend to grossly 
overfit the training data, and hence provide little, if any, actual ``learning.''
A random forest overcomes the weakness of decision trees by the use of
bagging (i.e., bootstrap aggregation), that is, multiple decision trees are
trained using subsets of the data and features. Then, a voting procedure 
based on these multiple decision trees is typically used to determine the 
random forest classification. In each of our random forest experiments,
we use~100 trees.
\item[Random trees] are a subclass of random forests, where the 
bagging only involves the classifiers, not the data. We would generally
expect better results from a random forest, but random trees will be 
more efficient. Our random trees results are based on a single tree.
\item[J48] is a specific implementation of the C4.5 algorithm~\cite{C4.5}, 
which is a popular method for constructing decision trees. We would generally
expect random trees to outperform decision trees while, as mentioned above, 
random forests should typically outperform random trees. 
However, decision trees are more efficient
than random trees, which are more efficient than random forests.
Thus, it is worth experimenting with all three of these tree-based algorithms
to determine the proper tradeoff between efficiency and accuracy. 
\item[Artificial neural network] (ANN) represents a large class of machine learning
techniques that attempt to (loosely) model the behavior of neurons and trained using
backpropagation~\cite{Stamp18deep}. While ANNs
are not a new concept, having first been proposed in the 1940s, they have found 
renewed interest in recent years as computing power has become sufficient
to effectively deal with ``deep'' neural network, i.e., networks that include
many hidden layers. Such deep networks have pushed machine learning 
to new heights. For our ANN experiments, we use two hidden layers,
with~10 neurons per layer, the rectified linear unit (ReLU) for the activation
functions on the hidden layers, and a sigmoid function for the output layer.
Training consists of~100 epochs, with the learning rate set at~$\alpha=0.001$.
\item[Support vector machine] (SVM) is a popular and effective machine 
learning technique. According to~\cite{BnC}, 
``SVMs are a rare example of a methodology where geometric
intuition, elegant mathematics, theoretical guarantees, and
practical algorithms meet.'' When training an SVM, we attempt to find a separating
hyperplane that maximizes the ``margin,'' i.e., the minimum distance between
the classes in the training set. A particularly nice feature of SVMs is that we can map
the input data to a higher dimensional feature space, where we are much more
likely to be able to separate the data. And, thanks to the so-called ``kernel trick,''
this mapping entails virtually no computational penalty. All of our SVM 
experiments are based on a linear kernel function with~$\varepsilon=0.001$.
\item[Logistic model tree] (LMT) can be viewed as a hybrid of decision trees and
logistic regression. That is, in an LMT, decision trees are constructed, with logistic regression
functions at the leaf nodes~\cite{LHF}. In our LMT experiments, we use
a minimum of~15 instances, where each node is considered for splitting.
\item[Boosting] is a general---and generally fairly straightforward---approach to building a 
strong model from a collection of (weak) models. In this paper, we employ the 
well-known adaptive boosting algorithm, AdaBoost~\cite{Stamp17Ada}. 
\item[Multinomial \naive\ Bayes] is a form of \naive\ Bayes where the 
underlying probability is assumed to satisfy a multinomial distribution.
In \naive\ Bayes, we make a strong independence assumption, which results in 
an extremely simply ``model'' that often performs surprisingly well in practice.
\end{description}

The static analysis can be done using the Java Bytecode extracted after
disassembling the {\tt apk} file. Also we can extract permissions from the manifest file.
In this paper, we take advantage of the static analysis using permissions of applications
and use them for detecting malware and also classify different malware families. The
effectiveness of these techniques is analyzed using multiple machine learning algorithms.

\subsection{Selected Related Work}

The paper~\cite{ref7} discusses a tool the authors refer to as 
Appopscopy, which implements a semantic language-based Android signature
detection strategy. In their research, general signatures are 
created for each malware family. Signature matching is achieved using inter-component 
call graphs based on control flow properties and the results are enhanced using
static taint analysis. The authors report an accuracy of~90\%\ on 
a malware dataset containing~1027 samples, with the accuracy for
individual families ranging from a high of~100\%\ to a low of~38\%.


In the research~\cite{ref8}, the authors analyze a tool called SCanDroid that
they have developed. This tool extracts features based on data flow.  
The work in~\cite{ref10} 
relies on $k$-nearest neighbor classification based on a variety of features,
include incoming and outgoing SMS and calls, device status, running processes, 
and more. This work claims that an accuracy of~93.75\%\ is achieved.
In the research~\cite{aung},
the authors propose a framework and test a variety of machine learning algorithms 
to analyze features based on Android events and permissions.
Experimental results from a dataset of some~500 malware samples yield
a maximum accuracy of~91.75\%\ for a random forest model.

In the paper~\cite{ref14}, the authors propose a dynamic analysis technique that 
is focused on the frequency of system and API calls. A large
number of machine learning techniques are tested on a dataset of about~4500 malicious
Android apps. The authors give accuracy results ranging
from~74.53\%\ to~95.96\%. Again, a random forest algorithm achieves
the best results.

The research~\cite{ref15} discusses a dynamic analysis tool, TaintDroid.
This sophisticated system analyzes network traffic to search for anomalous 
behavior---the research is in a similar vein as~\cite{ref7}, but with the emphasis
on efficiency. Another Android system call analysis technique is 
considered in~\cite{ref3}. 

Our work is perhaps most closely related to the research
in~\cite{NC3} and~\cite{NC4} which, in turn,
built on the groundbreaking work of~\cite{NC5} and~\cite{NC7},
as well as that in~\cite{ref9}.
In~\cite{NC5}, for example, an accuracy of~93.9\%\ is attained over a
dataset of~5600 malicious Android apps.

The paper~\cite{NC4} considers static and dynamic analysis of
Android malware based on permissions and API calls, respectively.
A robustness analysis is presented, and it is suggested that
malware writers can most likely defeat detectors that rely on permissions.
We provide a more careful analysis in this paper and find
that such is not the case.



\section{\uppercase{Experiments and Results}}\label{sect:exp}

\noindent In this section, we first discuss our datasets and feature extraction
process. Then we turn our attention to feature engineering, that is, we determine
the most significant features for use in our experiments. We also discuss
our experimental design before presenting results from
a wide variety experiments.

\subsection{Datasets}

We use the Android Malware
Genome Project~\cite{NC1} dataset. This data consists mainly of {\tt apk} files 
obtained from various malware forums and Android markets---these samples
have been widely used in previous research. Labels are included,
which specify the family to which each sample belongs. Thus, the data
can be used for both  binary classification (i.e., malware versus benign) and 
the multiclass (i.e., family) classification problems.

For our benign dataset, we crawled the PlayDrone project~\cite{PlayDrone},
as found on the Internet Archive~\cite{NC6}. The resulting {\tt apk} files
might include malicious samples. Therefore, we used
Androguard~\cite{NC2} to filter broken and potentially malicious {\tt apk} files. 
Table~\ref{tab:samps} gives the number of malware and benign samples that we
obtained. These samples will be used in our binary classification (malware versus benign)
and multiclass (malware family) experiments discussed below.

\begin{table}[htb]
\caption{Datasets}\label{tab:samps}
\centering
\begin{tabular}{ccc}\hline\hline
                    & \multicolumn{2}{c}{Samples} \\
Experiment & Type  & Number \\ \toprule
\multirow{2}{*}{Detection}    & Malware  & \z989 \\
                   & Benign & 2657 \\ \hline
Classification & Malware & 1260 \\ \hline\hline
\end{tabular}
\end{table}


\subsection{Feature Extraction}

To extract static features, we need to reverse engineer the {\tt apk} files. We 
again use Androguard this reverse engineering task. The manifest file,
{\tt AndroidManifest.xml}, contains numerous potential static features;
here we focus on the permissions requested by an application. 

\begin{figure}[!htb]
	\centering
	\includegraphics[width=0.425\textwidth]{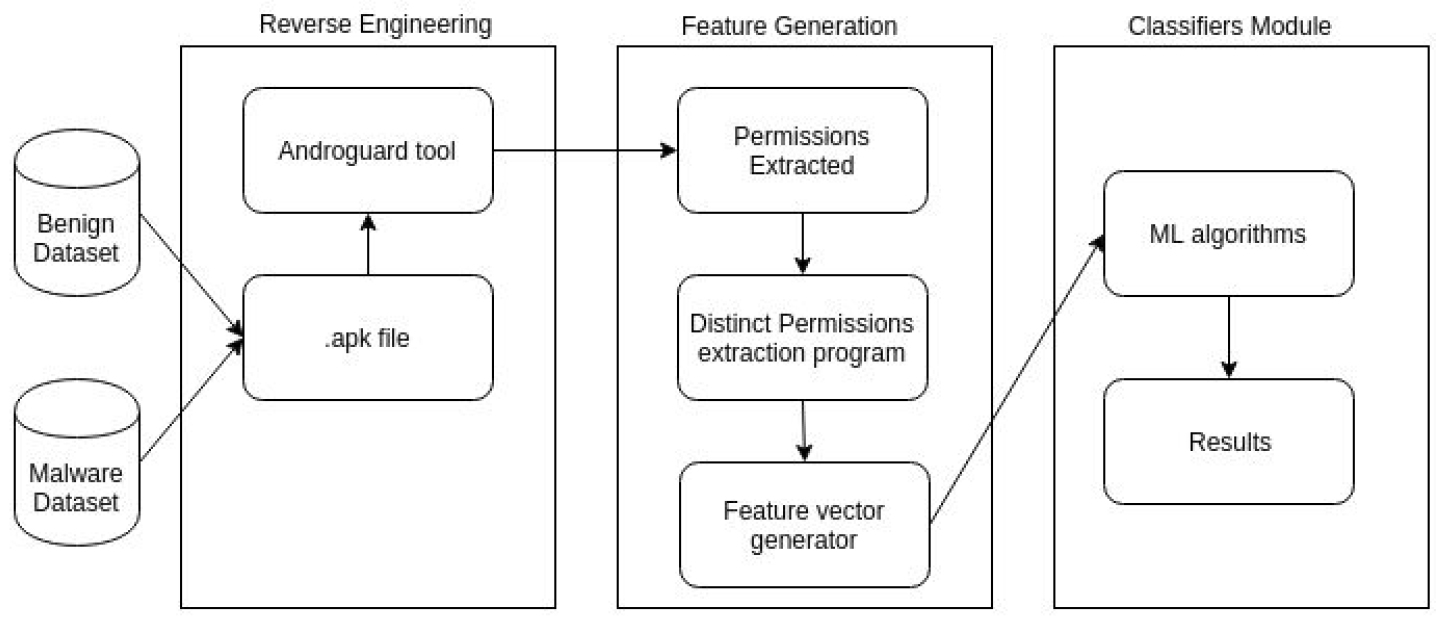}
	\caption{Binary classification architecture}\label{fig:1_mbca}
\end{figure}

From the superset of malware and benign samples, we find that there are~230 distinct
permissions. Thus, for each {\tt apk}, a feature vector is generated based on these
permissions. The feature vector is simply a binary sequence of length~230,
which indicates whether each of the corresponding permissions is 
requested by the application or not. Along with each feature vector,
we have a denoting label of~$+1$ or~$-1$, indicating whether the sample
is malware or benign, respectively. The overall architecture, in the case of binary
classification, is given in Figure~\ref{fig:1_mbca}.

For the multiclass (family) classification problem, essentially 
the same process is followed as for the binary classification case. 
However, we only examine malware samples, and over our malware
dataset, we find that only~118 distinct permissions occur. 
Thus, the feature vectors for the multiclass problem are of length~118.


\subsection{Feature Engineering}\label{sect:FE}

It is likely that many of the features under consideration
(i.e., permissions) provide little or no 
discriminating information, with respect to the malware versus benign
or the malware classification problem. It is useful to remove such features
from the analysis, as they essentially act as noise, and can therefore cause us
to obtain worse results than we would with a smaller, but more 
informative, feature set. It is also useful to remove extraneous features
so that scoring is as efficient as possible. Consequently, our immediate goal is to
determine features that are of no value for our analysis, and remove them
from subsequent consideration.


There are several techniques for determining feature significance. Here we
consider two distinct approaches to this problem. First, we use information gain
to reduce the feature set. Second, we use recursive feature elimination (RFE)
based on a linear SVM. Information gain is easily computed and gives us
a straightforward means of eliminating features. RFE is somewhat more involved,
but accounts for feature interactions in a way that a simple information gain
calculation cannot.

The information gain (IG)
provided by a feature is defined as the expected
reduction in entropy when we branch on that feature.
In the context of a decision tree, information gain
can be computed as the entropy of the parent node minus the average 
weighted entropy of its child nodes.
We measure the information gain for each feature, 
and select features in a greedy manner.
In a decision tree, this has the desirable effect of putting
decisions based on the most informative features closest
to the root. This is desirable, since the entropy is reduced
as rapidly as possible, and enables the tree to be simplified
by trimming features that provide little or no gain.

For our purposes, we simply use the information gain 
to reduce the number of features, then apply various 
machine learning techniques to this reduced feature set.
Based on the information gain, we selected the~74 highest ranked 
features---the top~10 of these features are given in Table~\ref{tab:IG_perm}.
Features that ranked outside the top~74 provided no
improvement in our results.

\begin{table}[htb]
\caption{Permissions ranked by IG}\label{tab:IG_perm}
\centering
\begin{tabular}{cl}\hline\hline
Score & Permission \\ \toprule
0.30682 & \fs READ\un SMS \\
0.28129 & \fs WRITE\un SMS \\
0.17211 & \fs READ\un PHONE\un STATE \\
0.15197 & \fs RECEIVE\un BOOT\un COMPLETED \\
0.14087 & \fs WRITE\un APN\un SETTINGS \\
0.13045 & \fs RECEIVE\un SMS \\
0.10695 & \fs SEND\un SMS \\
0.10614 & \fs CHANGE\un WIFI\un STATE \\
0.10042 & \fs INSTALL\un PACKAGES \\
0.10019 & \fs RESTART\un PACKAGES \\ \hline\hline
\end{tabular}
\end{table}


As mentioned above, we also reduce the feature set using RFE based on a linear SVM.
In a linear SVM, a weight is assigned to each feature, with the weight signifying the 
importance that the SVM attaches to the feature. For our RFE approach, we 
eliminate the feature with the lowest linear SVM weight, then train a new SVM
on this reduced (by one) feature set. Then we again eliminate the feature with
the lowest SVM weight, train a new linear SVM on this reduced feature set.
This process is continued until a single feature remains, and in this way, we
obtain a complete ranking of the features. The potential advantage of this RFE
technique is that it accounts for feature interactions among all of the reduced 
feature sets.
The top~10 features obtained using RFE based on a linear SVM
are listed in Table~\ref{tab:RFE_perm}.

\begin{table}[htb]
\caption{Permissions ranked by RFE using a linear SVM}\label{tab:RFE_perm}
\centering
\begin{tabular}{cl}\hline\hline
Rank & Permission \\ \toprule
\z1 & \fs WRITE\un APN\un SETTINGS \\
\z2 & \fs WRITE\un CALENDAR \\
\z3 & \fs WRITE\un CALL\un LOG \\
\z4 & \fs WRITE\un CONTACTS \\
\z5 & \fs WRITE\un INTERNAL\un STORAGE \\
\z6 & \fs WRITE\un OWNER\un DATA \\
\z7 & \fs WRITE\un SECURE\un SETTINGS \\
\z8 & \fs WRITE\un SETTINGS \\
\z9 & \fs WRITE\un SMS \\
10 & \fs WRITE\un SYNC\un SETTINGS \\ \hline\hline
\end{tabular}
\end{table}

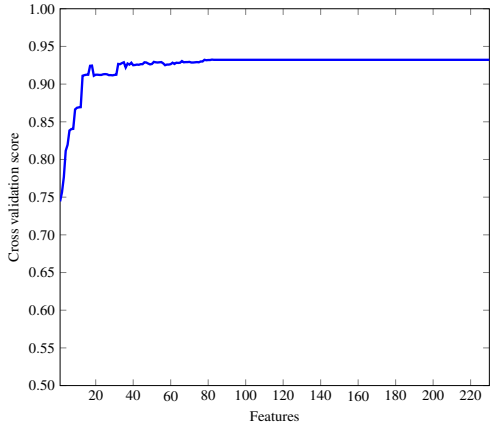
\begin{figure}[!htb]
	\centering
\begin{tikzpicture}[scale=0.55]
\begin{axis}[width=0.75\textwidth,
		   height=0.675\textwidth,
		   y tick label style={
			/pgf/number format/.cd,
    			fixed,
    			fixed zerofill,
    			precision=2
  			},
  		   x tick label style={
    			/pgf/number format/.cd,
    			fixed,
    			fixed zerofill,
    			precision=0
  		  	},
		   /pgf/number format/1000 sep={},
                    xmin=1,xmax=230,
                    ymin=0.5,ymax=1.0,
                    legend pos=south west,
                    xlabel={Features},
                    ylabel={Cross validation score}] 
\addplot[color=blue,ultra thick,mark=none] coordinates {
(1,0.744385)
(2,0.756439)
(3,0.776741)
(4,0.811536)
(5,0.819503)
(6,0.838433)
(7,0.840354)
(8,0.840628)
(9,0.866417)
(10,0.868606)
(11,0.869157)
(12,0.869431)
(13,0.911091)
(14,0.911916)
(15,0.912464)
(16,0.912465)
(17,0.924286)
(18,0.924561)
(19,0.911098)
(20,0.912471)
(21,0.912471)
(22,0.912195)
(23,0.912195)
(24,0.913020)
(25,0.913294)
(26,0.913020)
(27,0.911925)
(28,0.911925)
(29,0.911650)
(30,0.912198)
(31,0.912473)
(32,0.926747)
(33,0.926471)
(34,0.927841)
(35,0.928937)
(36,0.922087)
(37,0.927293)
(38,0.925924)
(39,0.928390)
(40,0.925104)
(41,0.925378)
(42,0.925926)
(43,0.925651)
(44,0.926474)
(45,0.926474)
(46,0.928940)
(47,0.928666)
(48,0.927295)
(49,0.926199)
(50,0.926748)
(51,0.929488)
(52,0.928942)
(53,0.928667)
(54,0.928941)
(55,0.929214)
(56,0.927570)
(57,0.925105)
(58,0.925928)
(59,0.925928)
(60,0.926476)
(61,0.928395)
(62,0.926750)
(63,0.928392)
(64,0.928118)
(65,0.928393)
(66,0.930585)
(67,0.928941)
(68,0.929215)
(69,0.929216)
(70,0.929490)
(71,0.928667)
(72,0.928667)
(73,0.928942)
(74,0.929216)
(75,0.928941)
(76,0.930039)
(77,0.930039)
(78,0.932231)
(79,0.931683)
(80,0.931957)
(81,0.931957)
(82,0.932505)
(83,0.932231)
(84,0.932231)
(85,0.932231)
(86,0.932231)
(87,0.932231)
(88,0.932231)
(89,0.932231)
(90,0.932231)
(91,0.932231)
(92,0.932231)
(93,0.932231)
(94,0.932231)
(95,0.932231)
(96,0.932231)
(97,0.932231)
(98,0.932231)
(99,0.932231)
(100,0.932231)
(101,0.932231)
(102,0.932231)
(103,0.932231)
(104,0.932231)
(105,0.932231)
(106,0.932231)
(107,0.932231)
(108,0.932231)
(109,0.932231)
(110,0.932231)
(111,0.932231)
(112,0.932231)
(113,0.932231)
(114,0.932231)
(115,0.932231)
(116,0.932231)
(117,0.932231)
(118,0.932231)
(119,0.932231)
(120,0.932231)
(121,0.932231)
(122,0.932231)
(123,0.932231)
(124,0.932231)
(125,0.932231)
(126,0.932231)
(127,0.932231)
(128,0.932231)
(129,0.932231)
(130,0.932231)
(131,0.932231)
(132,0.932231)
(133,0.932231)
(134,0.932231)
(135,0.932231)
(136,0.932231)
(137,0.932231)
(138,0.932231)
(139,0.932231)
(140,0.932231)
(141,0.932231)
(142,0.932231)
(143,0.932231)
(144,0.932231)
(145,0.932231)
(146,0.932231)
(147,0.932231)
(148,0.932231)
(149,0.932231)
(150,0.932231)
(151,0.932231)
(152,0.932231)
(153,0.932231)
(154,0.932231)
(155,0.932231)
(156,0.932231)
(157,0.932231)
(158,0.932231)
(159,0.932231)
(160,0.932231)
(161,0.932231)
(162,0.932231)
(163,0.932231)
(164,0.932231)
(165,0.932231)
(166,0.932231)
(167,0.932231)
(168,0.932231)
(169,0.932231)
(170,0.932231)
(171,0.932231)
(172,0.932231)
(173,0.932231)
(174,0.932231)
(175,0.932231)
(176,0.932231)
(177,0.932231)
(178,0.932231)
(179,0.932231)
(180,0.932231)
(181,0.932231)
(182,0.932231)
(183,0.932231)
(184,0.932231)
(185,0.932231)
(186,0.932231)
(187,0.932231)
(188,0.932231)
(189,0.932231)
(190,0.932231)
(191,0.932231)
(192,0.932231)
(193,0.932231)
(194,0.932231)
(195,0.932231)
(196,0.932231)
(197,0.932231)
(198,0.932231)
(199,0.932231)
(200,0.932231)
(201,0.932231)
(202,0.932231)
(203,0.932231)
(204,0.932231)
(205,0.932231)
(206,0.932231)
(207,0.932231)
(208,0.932231)
(209,0.932231)
(210,0.932231)
(211,0.932231)
(212,0.932231)
(213,0.932231)
(214,0.932231)
(215,0.932231)
(216,0.932231)
(217,0.932231)
(218,0.932231)
(219,0.932231)
(220,0.932231)
(221,0.932231)
(222,0.932231)
(223,0.932231)
(224,0.932231)
(225,0.932231)
(226,0.932231)
(227,0.932231)
(228,0.932231)
(229,0.932231)
(230,0.932231)
};
\end{axis}
\end{tikzpicture}
	\caption{Recursive feature elimination (RFE)}\label{fig:3_rfe}
\end{figure}

In Figure~\ref{fig:3_rfe}, we give the cross validation score of the linear SVM
as a function of the number of features, as obtained by RFE.
We see that the top~82 features gives us an optimal result---additional
features beyond this number provide no benefit. Consequently,
we use the~82 top RFE features in our experiments below.

\subsection{Binary Classification}

In this section, we discuss our binary classification experiments.
We consider both the IG features and the RFE features. We
test a wide variety of machine learning algorithms and provide 
various statistics.

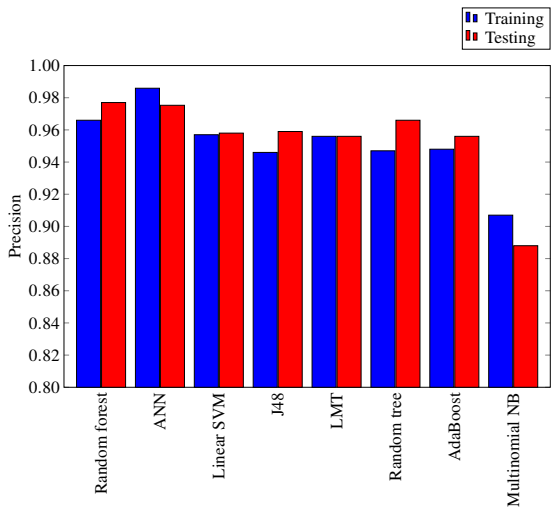
\begin{figure}[!htb]
	\centering
\begin{tikzpicture}[scale=0.45]
    \begin{axis}[
        ymin=0.8,
	ymax=1.0,
        width  = 1.0*\textwidth,
        height = 0.7*\textwidth,
        major x tick style = transparent,
        ybar=2*\pgflinewidth,
        bar width=20pt,
        ylabel = {Precision},
        symbolic x coords={Random forest,ANN,Linear SVM,J48,LMT,Random tree,AdaBoost,Multinomial NB},
	y label style={font=\LARGE},
	y tick label style={
	  	font=\LARGE,
  	  	/pgf/number format/.cd,
   		fixed,
   	  	fixed zerofill,
    	  	precision=2},
        x tick label style={
        		rotate=90,
		anchor=east,
		inner sep=0mm
		},
        scaled y ticks = false,
        every tick label/.append style={font=\LARGE},
        enlarge x limits=0.09,
        legend cell align=left,
        legend style={
                at={(1,1.05)},
                anchor=south east,
                font=\LARGE,
                column sep=1ex
        }
    ]
        \addplot[fill=blue,opacity=1.0]
            coordinates {
(Random forest,0.966)
(ANN,0.9859)
(Linear SVM,0.957)
(J48,0.946)
(LMT,0.956)
(Random tree,0.947)
(AdaBoost,0.948)
(Multinomial NB,0.907)
};
        \addplot[fill=red,opacity=1.0]
            coordinates {
(Random forest,0.977)
(ANN,0.9753)
(Linear SVM,0.958)
(J48,0.959)
(LMT,0.956)
(Random tree,0.966)
(AdaBoost,0.956)
(Multinomial NB,0.888)
};
\legend{Training,Testing}
    \end{axis}
\end{tikzpicture}
	\caption{Machine learning comparison based on precision (IG features)}\label{fig:4_IG_prec}
\end{figure}

First, we consider the IG features, as discussed in Section~\ref{sect:FE}, above.
Bar graphs of training and testing precision for each of the techniques 
discussed in Section~\ref{sect:ML_techs}
are plotted in Figure~\ref{fig:4_IG_prec}. For this experiment the 
number of samples in the malware and benign datasets are listed in Table~\ref{tab:samps},
where we see that there is total of~3647 samples. Here, and in all subsequent experiments,
we use 5-fold cross validation. Cross validation serves to maximize the number
of test cases, while simultaneously minimizing the effect of any 
bias that might exist in the training data~\cite{StampML2017}.


\begin{figure}[!htb]
	\centering
\begin{tikzpicture}[scale=0.45]
    \begin{axis}[
        ymin=0.8,
	ymax=1.0,
        width  = 1.0*\textwidth,
        height = 0.7*\textwidth,
        major x tick style = transparent,
        ybar=2*\pgflinewidth,
        bar width=20pt,
        ylabel = {AUC},
        symbolic x coords={Random forest,ANN,Linear SVM,J48,LMT,Random tree,AdaBoost,Multinomial NB},
	y label style={font=\LARGE},
	y tick label style={
	  	font=\LARGE,
  	  	/pgf/number format/.cd,
   		fixed,
   	  	fixed zerofill,
    	  	precision=2},
        x tick label style={
        		rotate=90,
		anchor=east,
		inner sep=0mm
		},
        scaled y ticks = false,
        every tick label/.append style={font=\LARGE},
        enlarge x limits=0.09,
        legend cell align=left,
        legend style={
                at={(1,1.05)},
                anchor=south east,
                font=\LARGE,
                column sep=1ex
        }
    ]
        \addplot[fill=blue,opacity=1.0]
            coordinates {
(Random forest,0.986)
(ANN,0.9946)
(Linear SVM,0.938)
(J48,0.946)
(LMT,0.971)
(Random tree,0.941)
(AdaBoost,0.980)
(Multinomial NB,0.948)
};
        \addplot[fill=red,opacity=1.0]
            coordinates {
(Random forest,0.987)
(ANN,0.9662)
(Linear SVM,0.941)
(J48,0.972)
(LMT,0.969)
(Random tree,0.967)
(AdaBoost,0.988)
(Multinomial NB,0.939)
};
\legend{Training,Testing}
    \end{axis}
\end{tikzpicture}
	\caption{Machine learning comparison based on AUC  (IG features)}\label{fig:5_IG_AUC}
\end{figure}
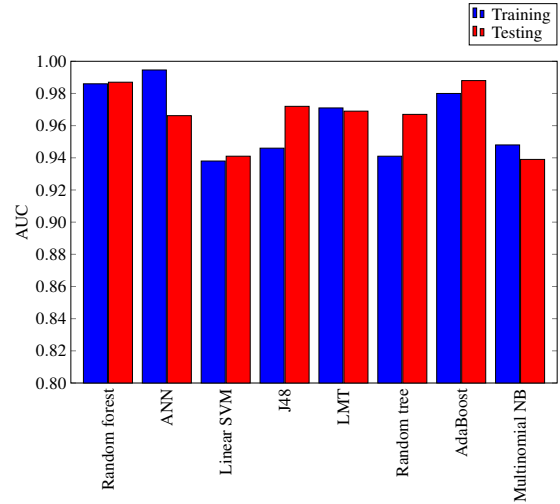

Given a scatterplot of experimental results, an ROC curve is obtained by 
graphing the true positive rate versus the false positive rate, as the
threshold varies through the range of values. The area under the ROC curve (AUC)
is between~0 and~1, inclusive, and can be interpreted as the probability that a randomly
selected positive instance scores higher than a randomly selected negative 
instance~\cite{StampML2017}.
In Figure~\ref{fig:5_IG_AUC}, we give the AUC statistic for the same set of IG feature
experiments that we ahve summarized in Figure~\ref{fig:4_IG_prec}.

\begin{figure}[!htb]
	\centering
\begin{tikzpicture}[scale=0.45]
    \begin{axis}[
        ymin=0.8,
	ymax=1.0,
        width  = 1.0*\textwidth,
        height = 0.7*\textwidth,
        major x tick style = transparent,
        ybar=2*\pgflinewidth,
        bar width=20pt,
        ylabel = {Precision},
        symbolic x coords={Random forest,ANN,Linear SVM,J48,LMT,Random tree,AdaBoost,Multinomial NB},
	y label style={font=\LARGE},
	y tick label style={
	  	font=\LARGE,
  	  	/pgf/number format/.cd,
   		fixed,
   	  	fixed zerofill,
    	  	precision=2},
        x tick label style={
        		rotate=90,
		anchor=east,
		inner sep=0mm
		},
        scaled y ticks = false,
        every tick label/.append style={font=\LARGE},
        enlarge x limits=0.09,
        legend cell align=left,
        legend style={
                at={(1,1.05)},
                anchor=south east,
                font=\LARGE,
                column sep=1ex
        }
    ]
        \addplot[fill=blue,opacity=1.0]
            coordinates {
(Random forest,0.966)
(ANN,0.9815)
(Linear SVM,0.961)
(J48,0.952)
(LMT,0.958)
(Random tree,0.951)
(AdaBoost,0.952)
(Multinomial NB,0.912)
};
        \addplot[fill=red,opacity=1.0]
            coordinates {
(Random forest,0.970)
(ANN,0.960)
(Linear SVM,0.959)
(J48,0.958)
(LMT,0.962)
(Random tree,0.963)
(AdaBoost,0.964)
(Multinomial NB,0.904)
};
\legend{Training,Testing}
    \end{axis}
\end{tikzpicture}
	\caption{Machine learning comparison based on precision  (RFE features)}\label{fig:6_RFE_prec}
\end{figure}
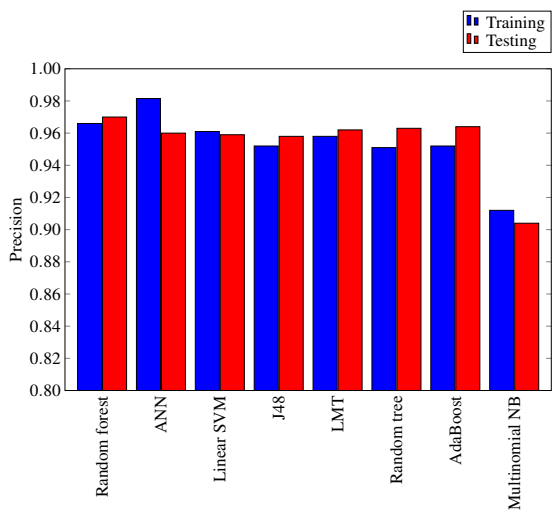

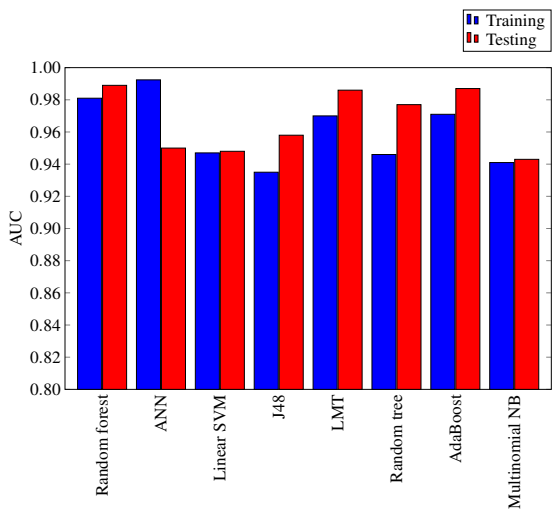
\begin{figure}[!htb]
	\centering
\begin{tikzpicture}[scale=0.45]
    \begin{axis}[
        ymin=0.8,
	ymax=1.0,
        width  = 1.0*\textwidth,
        height = 0.7*\textwidth,
        major x tick style = transparent,
        ybar=2*\pgflinewidth,
        bar width=20pt,
        ylabel = {AUC},
        symbolic x coords={Random forest,ANN,Linear SVM,J48,LMT,Random tree,AdaBoost,Multinomial NB},
	y label style={font=\LARGE},
	y tick label style={
	  	font=\LARGE,
  	  	/pgf/number format/.cd,
   		fixed,
   	  	fixed zerofill,
    	  	precision=2},
        x tick label style={
        		rotate=90,
		anchor=east,
		inner sep=0mm
		},
        scaled y ticks = false,
        every tick label/.append style={font=\LARGE},
        enlarge x limits=0.09,
        legend cell align=left,
        legend style={
                at={(1,1.05)},
                anchor=south east,
                font=\LARGE,
                column sep=1ex
        }
    ]
        \addplot[fill=blue,opacity=1.0]
            coordinates {
(Random forest,0.981)
(ANN,0.9924)
(Linear SVM,0.947)
(J48,0.935)
(LMT,0.970)
(Random tree,0.946)
(AdaBoost,0.971)
(Multinomial NB,0.941)
};
        \addplot[fill=red,opacity=1.0]
            coordinates {
(Random forest,0.989)
(ANN,0.95)
(Linear SVM,0.948)
(J48,0.958)
(LMT,0.986)
(Random tree,0.977)
(AdaBoost,0.987)
(Multinomial NB,0.943)
};
\legend{Training,Testing}
    \end{axis}
\end{tikzpicture}
	\caption{Machine learning comparison based on AUC (RFE features)}\label{fig:7_RFE_AUC}
\end{figure}

We repeated the experiments above using the~82 RFE features, rather than
the~74 IG features. The precision results for these machine learning experiments
are given in Figure~\ref{fig:6_RFE_prec}, while the corresponding AUC results are 
summarized in Figure~\ref{fig:7_RFE_AUC}.

The performance between the various machine learning algorithms varies
significantly, with multinomial \naive\ Bayes consistently the worst,
while random forests and ANNs perform the best. The IG and RFE
cases are fairly similar, although ANNs are better on the RFE
features, with random forest are better on the IG features.

\subsection{ANN Experiments}\label{sect:ANN}

Since ANNs performed well in the experiments above, we have
conducted additional experiments to determine the effect of an imbalanced
dataset and to test the effect of small training sets.
For these experiments, we use the IG features
and the same binary classification dataset as above, with the skewed
training sets selected at random.
Again, we use 5-fold cross validation.

We experiment with three different ratios between the sizes of the
malware and benign sets, namely, a ratio of~1:3 (i.e., three
times as many benign samples as malware samples), as well as 
ratios of~1:6 and~1:12. For the~1:3 ratio, we have sufficient data to consider 
following four different cases:
\begin{itemize}
\item 100 malware and 300 benign
\item 200 malware and 600 benign
\item 400 malware and 1200 benign
\item 800 malware and 2400 benign
\end{itemize}
For a~1:6 ratio, we have enough samples so that we can 
consider the following three cases:
\begin{itemize}
\item 100 malware and 600 benign
\item 200 malware and 1200 benign
\item 400 malware and 2400 benign
\end{itemize}
Finally, for the~1:12 ratio, we have sufficient data
for the following two cases:
\begin{itemize}
\item 100 malware and 1200 benign
\item 200 malware and 2400 benign
\end{itemize}

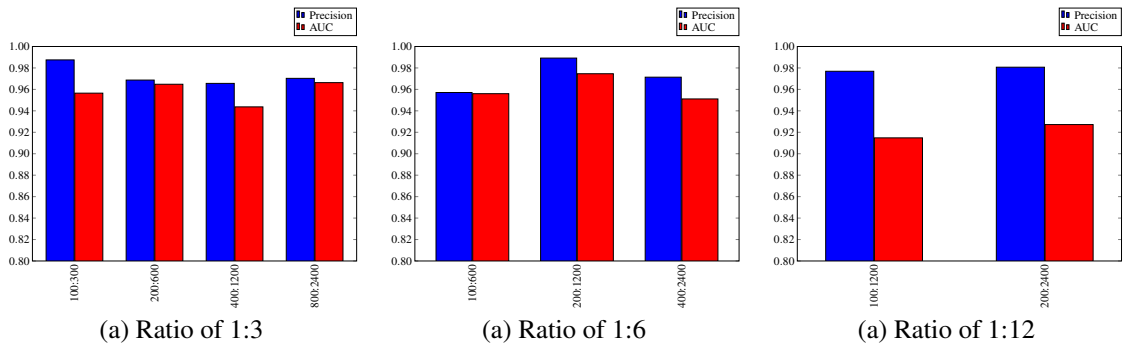
\begin{figure*}[!htb]
	\centering
	\begin{tabular}{ccc}
\begin{tikzpicture}[scale=0.30]
    \begin{axis}[
        ymin=0.8,
	ymax=1.0,
        width  = 1.0*\textwidth,
        height = 0.7*\textwidth,
        major x tick style = transparent,
        ybar=2*\pgflinewidth,
        bar width=35pt,
        symbolic x coords={100:300,200:600,400:1200,800:2400},
	y label style={font=\LARGE},
	y tick label style={
	  	font=\LARGE,
  	  	/pgf/number format/.cd,
   		fixed,
   	  	fixed zerofill,
    	  	precision=2},
         xtick={100:300,200:600,400:1200,800:2400},
        x tick label style={
        		rotate=90,
		anchor=east,
		inner sep=0mm
		},
        scaled y ticks = false,
        every tick label/.append style={font=\LARGE},
        enlarge x limits=0.175,
        legend cell align=left,
        legend style={
                at={(1,1.05)},
                anchor=south east,
                font=\LARGE,
                column sep=1ex
        }
    ]
        \addplot[fill=blue,opacity=1.0]
            coordinates {
(100:300,0.9875)
(200:600,0.9687)
(400:1200,0.9656)
(800:2400,0.9703)
};
        \addplot[fill=red,opacity=1.0]
            coordinates {
(100:300,0.9565)
(200:600,0.9648)
(400:1200,0.9437)
(800:2400,0.9663)
};
\legend{Precision,AUC}
    \end{axis}
\end{tikzpicture}
	&
\begin{tikzpicture}[scale=0.30]
    \begin{axis}[
        ymin=0.8,
	ymax=1.0,
        width  = 1.0*\textwidth,
        height = 0.7*\textwidth,
        major x tick style = transparent,
        ybar=2*\pgflinewidth,
        bar width=45pt,
        symbolic x coords={100:600,200:1200,400:2400},
	y label style={font=\LARGE},
	y tick label style={
	  	font=\LARGE,
  	  	/pgf/number format/.cd,
   		fixed,
   	  	fixed zerofill,
    	  	precision=2},
         xtick={100:600,200:1200,400:2400},
        x tick label style={
        		rotate=90,
		anchor=east,
		inner sep=0mm
		},
        scaled y ticks = false,
        every tick label/.append style={font=\LARGE},
        enlarge x limits=0.275,
        legend cell align=left,
        legend style={
                at={(1,1.05)},
                anchor=south east,
                font=\LARGE,
                column sep=1ex
        }
    ]
        \addplot[fill=blue,opacity=1.0]
            coordinates {
(100:600,0.9571)
(200:1200,0.9892)
(400:2400,0.9714)
};
        \addplot[fill=red,opacity=1.0]
            coordinates {
(100:600,0.9560)
(200:1200,0.9746)
(400:2400,0.9511)
};
\legend{Precision,AUC}
    \end{axis}
\end{tikzpicture}
	&
\begin{tikzpicture}[scale=0.30]
    \begin{axis}[
        ymin=0.8,
	ymax=1.0,
        width  = 1.0*\textwidth,
        height = 0.7*\textwidth,
        major x tick style = transparent,
        ybar=2*\pgflinewidth,
        bar width=60pt,
        symbolic x coords={100:1200,200:2400},
	y label style={font=\LARGE},
	y tick label style={
	  	font=\LARGE,
  	  	/pgf/number format/.cd,
   		fixed,
   	  	fixed zerofill,
    	  	precision=2},
         xtick={100:1200,200:2400},
        x tick label style={
        		rotate=90,
		anchor=east,
		inner sep=0mm
		},
        scaled y ticks = false,
        every tick label/.append style={font=\LARGE},
        enlarge x limits=0.45,
        legend cell align=left,
        legend style={
                at={(1,1.05)},
                anchor=south east,
                font=\LARGE,
                column sep=1ex
        }
    ]
        \addplot[fill=blue,opacity=1.0]
            coordinates {
(100:1200,0.9769)
(200:2400,0.9807)
};
        \addplot[fill=red,opacity=1.0]
            coordinates {
(100:1200,0.9148)
(200:2400,0.9272)
};
\legend{Precision,AUC}
    \end{axis}
\end{tikzpicture}
	\\
	(a) Ratio of 1:3
	&
	(a) Ratio of 1:6
	&
	(a) Ratio of 1:12
	\\[2ex]
	\end{tabular}
	\caption{ANN results for various malware to benign ratios}\label{fig:8_ANN}
\end{figure*}

The testing precision and AUC results for 
the~1:3, 1:6, and~1:12 training cases are given in
Figures~\ref{fig:8_ANN}~(a) through~(c), respectively.
We see that,
for example, with only~100 malware and~300 benign samples,
we obtain a testing precision in excess of~0.98 and
an AUC of approximately~0.96. Overall, these results show
that the ANN performs extremely well, even with
a small and highly skewed training set. This is significant,
since we would like to train a model as soon as possible
(in which case the training set may be small), and the
samples are likely to be highly skewed towards the benign case.

\subsection{Robustness Experiments}

As an attack on permissions-based detection, a malware writer might simply
request more of the permissions that are typical of benign samples, while still requesting
permissions that are necessary for the malware to function. In this way, the
permissions statistics of the malware samples would be somewhat 
closer to those of the benign samples. 
Analogous attacks have proven successful against 
malware detection based on opcode sequences~\cite{LinS11}.

\begin{figure}[!htb]
	\begin{tikzpicture}[scale=0.925]
\begin{axis}[
		   width=0.45\textwidth,
		   height=0.4\textwidth,
		   /pgf/number format/1000 sep={},
                    xmin=0,xmax=20,
                    ymin=0.8,ymax=1.0,
                    legend pos=south east,
                    xticklabel style={
                    	/pgf/number format/fixed,
			/pgf/number format/precision=0},
                    yticklabel style={
                    	/pgf/number format/fixed,
			/pgf/number format/precision=2},
                    xlabel={$N$},
                    ] 
\addplot[color=blue,ultra thick,mark=star,mark size=2.0] coordinates {
(0,0.98)
(1,0.944)
(2,0.935)
(3,0.95)
(4,0.93)
(5,0.93)
(6,0.981)
(7,0.986)
(8,0.986)
(9,0.986)
(10,1.0)
(11,0.995)
(12,0.995)
(13,1.0)
(14,0.995)
(15,0.990)
(16,0.995)
(17,0.995)
(18,0.995)
(19,0.995)
(20,0.995)
};
\addlegendentry{Precision}
\addplot[color=red,ultra thick,mark=star,mark size=2.0] coordinates {
(0,0.97)
(1,0.957)
(2,0.956)
(3,0.958)
(4,0.954)
(5,0.957)
(6,0.982)
(7,0.988)
(8,0.989)
(9,0.989)
(10,0.995)
(11,0.989)
(12,0.993)
(13,0.996)
(14,0.993)
(15,0.995)
(16,0.998)
(17,0.997)
(18,0.997)
(19,0.997)
(20,0.997)
};
\addlegendentry{AUC}
\end{axis}
\end{tikzpicture}
	\caption{Robustness tests based on ANN}\label{fig:8a_robust}
\end{figure}
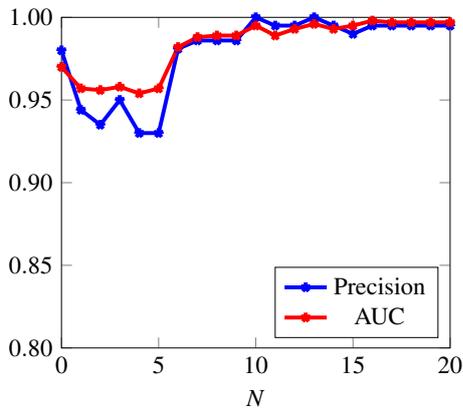

To simulate such an attack, for each value 
of~$N=0,1,2,\ldots,20$, we include the top~$N$ benign
permissions in each malware sample. For each of these cases, we have 
performed an ANN experiment, similar to those
in Section~\ref{sect:ANN}, above. 
The precision and AUC results are given in the form of line graphs in 
Figure~\ref{fig:8a_robust}. Note that the~$N=0$ case corresponds to 
no modification to the malware permissions.

The results in Figure~\ref{fig:8a_robust} show that there is a decrease 
in the effectiveness of the ANN when a small number of the most popular 
benign permissions are requested. However, when more than~$N=5$ permissions
are included, the success of the ANN recovers, and actually improves
on the unmodified~$N=0$ case. 
These results show that 
a straightforward attack on the permissions can have a modest effect, 
but we also see that permissions are a surprisingly robust feature.

\subsection{Multiclass Classification}

For the multiclass experiments in this section, we again use 
permission-based features.
The metrics considered here are precision 
and the AUC. As above, we use five-fold cross validation in each experiment.

\begin{figure*}[!htb]
	\centering
\begin{tikzpicture}[scale=1.0]
\begin{axis}[
        width  = 1.0*\textwidth,
        height = 7cm,
        major x tick style = transparent,
        ybar=2*\pgflinewidth,
        bar width=6.0pt,
        ylabel = {Number of samples},
        symbolic x coords={
FakeNetflix,
SndApps,
DroidKungFu4,
BeanBot,
RogueLemon,
Gone60,
SMSReplicator,
DroidKungFu1,
Zsone,
AnserverBot,
HippoSMS,
GingerMaster,
Pjapps,
Walkinwat,
DogWars,
GPSSMSSpy,
GGTracker,
FakePlayer,
Asroot,
Bgserv,
ADRD,
DroidKungFu3,
DroidKungFuUpdate,
KMin,
Spitmo,
Tapsnake,
CruseWin,
BaseBridge,
Endofday,
YZHC,
DroidKungFu2,
Jifake,
DroidKungFuSapp,
Geinimi,
GoldDream,
zHash,
DroidDeluxe,
LoveTrap,
DroidDream,
GamblerSMS,
Zitmo,
NickyBot,
NickySpy,
CoinPirate,
DroidCoupon,
RogueSPPush,
Plankton,
DroidDreamLight,
jSMSHider
},
	y tick label style={
    		/pgf/number format/.cd,
   		fixed,
   		fixed zerofill,
    		precision=0},
	ymin=0,
	ymax=355,
        nodes near coords,
        every node near coord/.append style={
		rotate=90,
		anchor=west,
        		font=\scriptsize,
		/pgf/number format/.cd,
   		fixed,
   		fixed zerofill,
    		precision=0},
        every tick label/.append style={font=\scriptsize},
        xtick = data,
        nodes near coords,
        x tick label style={
        		rotate=90,
		anchor=east, 
		inner sep=0mm},
        scaled y ticks = false,
        enlarge x limits=0.02,
        ymin=0,
        legend cell align=left,
        legend style={
                at={(1,1.05)},
                anchor=south east,
                column sep=1ex
        }
    ]
        \addplot[fill=blue,opacity=1.0]
            coordinates {
(FakeNetflix,1)
(SndApps,10)
(DroidKungFu4,96)
(BeanBot,8)
(RogueLemon,2)
(Gone60,9)
(SMSReplicator,1)
(DroidKungFu1,34)
(Zsone,12)
(AnserverBot,187)
(HippoSMS,4)
(GingerMaster,4)
(Pjapps,58)
(Walkinwat,1)
(DogWars,1)
(GPSSMSSpy,6)
(GGTracker,1)
(FakePlayer,6)
(Asroot,8)
(Bgserv,9)
(ADRD,22)
(DroidKungFu3,309)
(DroidKungFuUpdate,1)
(KMin,52)
(Spitmo,1)
(Tapsnake,2)
(CruseWin,2)
(BaseBridge,122)
(Endofday,1)
(YZHC,22)
(DroidKungFu2,30)
(Jifake,1)
(DroidKungFuSapp,3)
(Geinimi,69)
(GoldDream,47)
(zHash,11)
(DroidDeluxe,1)
(LoveTrap,1)
(DroidDream,16)
(GamblerSMS,1)
(Zitmo,1)
(NickyBot,1)
(NickySpy,2)
(CoinPirate,1)
(DroidCoupon,1)
(RogueSPPush,9)
(Plankton,11)
(DroidDreamLight,46)
(jSMSHider,16)
};
\end{axis}
\end{tikzpicture}
	\caption{Distribution of malware families}\label{fig:10_fam_dist}
\end{figure*}
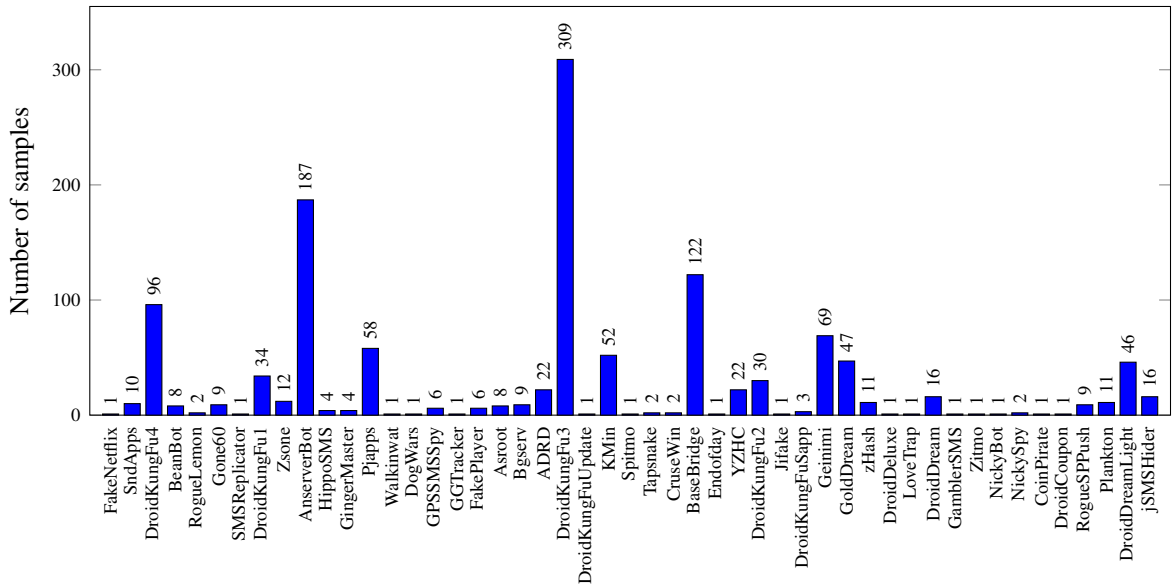

For these experiments, we use all malware samples in our dataset
that include a family label. The distribution of these malware families 
is given in Figure~\ref{fig:10_fam_dist}.  Note that we have~49 
malware families and a total~1260 malware samples.


%
%
%

\begin{figure*}[!htb]
	\centering
	\begin{tabular}{cc}
\begin{tikzpicture}[scale=0.45]
    \begin{axis}[
        ymin=0.7,
	ymax=1.0,
        width  = 1.0*\textwidth,
        height = 0.7*\textwidth,
        major x tick style = transparent,
        ybar=2*\pgflinewidth,
        bar width=27.5pt,
        ylabel = {Precision},
        symbolic x coords={J48,Random forest,LMT,Linear SVM,AdaBoost},
	y label style={font=\LARGE},
	y tick label style={
	  	font=\LARGE,
  	  	/pgf/number format/.cd,
   		fixed,
   	  	fixed zerofill,
    	  	precision=2},
         xtick={J48,Random forest,LMT,Linear SVM,AdaBoost},
        x tick label style={
        		rotate=90,
		anchor=east,
		inner sep=0mm
		},
        scaled y ticks = false,
        every tick label/.append style={font=\LARGE},
        enlarge x limits=0.125,
        legend cell align=left,
        legend style={
                at={(1,1.05)},
                anchor=south east,
                font=\LARGE,
                column sep=1ex
        }
    ]
        \addplot[fill=blue,opacity=1.0]
            coordinates {
(J48,0.923)
(Random forest,0.947)
(LMT,0.933)
(Linear SVM,0.940)
(AdaBoost,0.90)
};
        \addplot[fill=red,opacity=1.0]
            coordinates {
(J48,0.917)
(Random forest,0.944)
(LMT,0.930)
(Linear SVM,0.929)
(AdaBoost,0.90)
};
\legend{Training,Testing}
    \end{axis}
\end{tikzpicture}
	&
\begin{tikzpicture}[scale=0.45]
    \begin{axis}[
        ymin=0.7,
	ymax=1.0,
        width  = 1.0*\textwidth,
        height = 0.7*\textwidth,
        major x tick style = transparent,
        ybar=2*\pgflinewidth,
        bar width=27.5pt,
        ylabel = {AUC},
        symbolic x coords={J48,Random forest,LMT,Linear SVM,AdaBoost},
	y label style={font=\LARGE},
	y tick label style={
	  	font=\LARGE,
  	  	/pgf/number format/.cd,
   		fixed,
   	  	fixed zerofill,
    	  	precision=2},
         xtick={J48,Random forest,LMT,Linear SVM,AdaBoost},
        x tick label style={
        		rotate=90,
		anchor=east,
		inner sep=0mm
		},
        scaled y ticks = false,
        every tick label/.append style={font=\LARGE},
        enlarge x limits=0.125,
        legend cell align=left,
        legend style={
                at={(1,1.05)},
                anchor=south east,
                font=\LARGE,
                column sep=1ex
        }
    ]
        \addplot[fill=blue,opacity=1.0]
            coordinates {
(J48,0.975)
(Random forest,0.991)
(LMT,0.988)
(Linear SVM,0.968)
(AdaBoost,0.991)
};
        \addplot[fill=red,opacity=1.0]
            coordinates {
(J48,0.966)
(Random forest,0.987)
(LMT,0.990)
(Linear SVM,0.961)
(AdaBoost,0.990)
};
\legend{Training,Testing}
    \end{axis}
\end{tikzpicture}
	\\
	(a) Precision
	&
	(b) AUC
	\\[2ex]
	\end{tabular}
	\caption{Multiclass results}\label{fig:13_reduce}
\end{figure*}
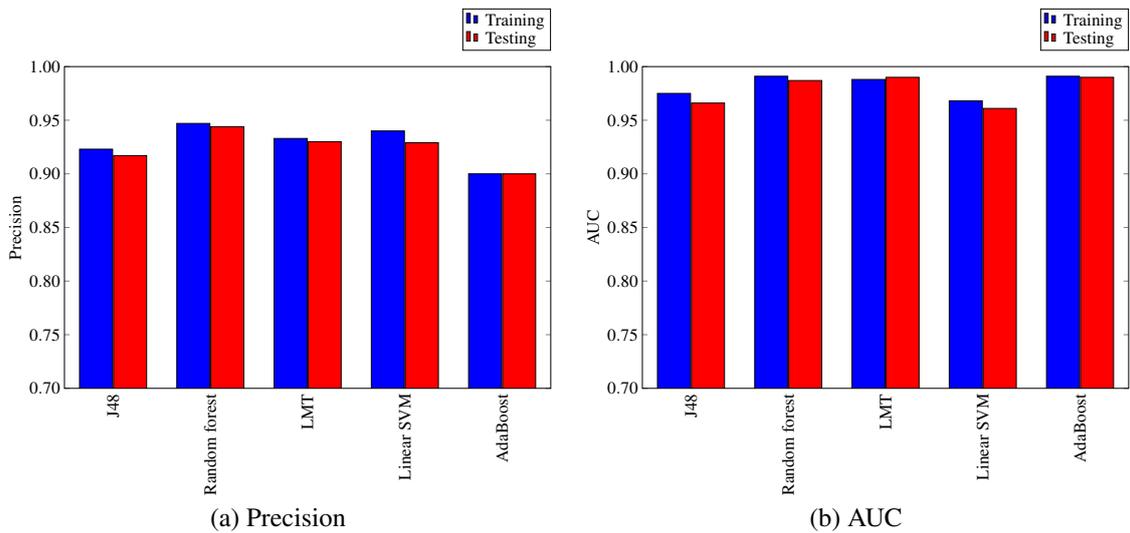

Bar graphs of precision and AUC results for multiple machine learning techniques 
are give in Figures~\ref{fig:13_reduce}~(a) and~(b),
respectively. We observe that a random forest performs best
in these multiclass experiments, attaining a testing accuracy 
in excess of~0.95. This is an impressive number, given that
we have such a large number of classes. Other machine learning
techniques that perform well on this challenging multiclass problem
include LMT and AdaBoost.

\begin{figure*}[!htb]
	\centering
	\begin{tabular}{ccccc}
\begin{tikzpicture}[scale=0.30]
    \begin{axis}[
        ymin=0.8,
	ymax=1.0,
        width  = 1.0*\textwidth,
        height = 0.7*\textwidth,
        major x tick style = transparent,
        ybar=2*\pgflinewidth,
        bar width=45pt,
        symbolic x coords={Random forest,LMT,Linear SVM},
	y label style={font=\LARGE},
	y tick label style={
	  	font=\LARGE,
  	  	/pgf/number format/.cd,
   		fixed,
   	  	fixed zerofill,
    	  	precision=2},
         xtick={Random forest,LMT,Linear SVM},
        x tick label style={
        		rotate=90,
		anchor=east,
		inner sep=0mm
		},
        scaled y ticks = false,
        every tick label/.append style={font=\LARGE},
        enlarge x limits=0.275,
        legend cell align=left,
        legend style={
                at={(1,1.05)},
                anchor=south east,
                font=\LARGE,
                column sep=1ex
        }
    ]
        \addplot[fill=blue,opacity=1.0]
            coordinates {
(Random forest,0.9)
(LMT,0.903)
(Linear SVM,0.929)
};
        \addplot[fill=red,opacity=1.0]
            coordinates {
(Random forest,0.996)
(LMT,0.994)
(Linear SVM,0.958)
};
\legend{Precision,AUC}
    \end{axis}
\end{tikzpicture}
	&
\begin{tikzpicture}[scale=0.30]
    \begin{axis}[
        ymin=0.8,
	ymax=1.0,
        width  = 1.0*\textwidth,
        height = 0.7*\textwidth,
        major x tick style = transparent,
        ybar=2*\pgflinewidth,
        bar width=45pt,
        symbolic x coords={Random forest,LMT,Linear SVM},
	y label style={font=\LARGE},
	y tick label style={
	  	font=\LARGE,
  	  	/pgf/number format/.cd,
   		fixed,
   	  	fixed zerofill,
    	  	precision=2},
         xtick={Random forest,LMT,Linear SVM},
        x tick label style={
        		rotate=90,
		anchor=east,
		inner sep=0mm
		},
        scaled y ticks = false,
        every tick label/.append style={font=\LARGE},
        enlarge x limits=0.275,
        legend cell align=left,
        legend style={
                at={(1,1.05)},
                anchor=south east,
                font=\LARGE,
                column sep=1ex
        }
    ]
        \addplot[fill=blue,opacity=1.0]
            coordinates {
(Random forest,0.965)
(LMT,0.965)
(Linear SVM,0.943)
};
        \addplot[fill=red,opacity=1.0]
            coordinates {
(Random forest,1.0)
(LMT,1.0)
(Linear SVM,0.961)
};
\legend{Precision,AUC}
    \end{axis}
\end{tikzpicture}
	&
\begin{tikzpicture}[scale=0.30]
    \begin{axis}[
        ymin=0.8,
	ymax=1.0,
        width  = 1.0*\textwidth,
        height = 0.7*\textwidth,
        major x tick style = transparent,
        ybar=2*\pgflinewidth,
        bar width=45pt,
        symbolic x coords={Random forest,LMT,Linear SVM},
	y label style={font=\LARGE},
	y tick label style={
	  	font=\LARGE,
  	  	/pgf/number format/.cd,
   		fixed,
   	  	fixed zerofill,
    	  	precision=2},
         xtick={Random forest,LMT,Linear SVM},
        x tick label style={
        		rotate=90,
		anchor=east,
		inner sep=0mm
		},
        scaled y ticks = false,
        every tick label/.append style={font=\LARGE},
        enlarge x limits=0.275,
        legend cell align=left,
        legend style={
                at={(1,1.05)},
                anchor=south east,
                font=\LARGE,
                column sep=1ex
        }
    ]
        \addplot[fill=blue,opacity=1.0]
            coordinates {
(Random forest,1.0)
(LMT,0.967)
(Linear SVM,1.0)
};
        \addplot[fill=red,opacity=1.0]
            coordinates {
(Random forest,1.0)
(LMT,1.0)
(Linear SVM,1.0)
};
\legend{Precision,AUC}
    \end{axis}
\end{tikzpicture}
	\\
	(a) 30 samples (11 families)
	&
	(b) 40 samples (9 families)
	&
	(c) 50 samples (7 families)
	\\[2ex]
	\end{tabular}
	\caption{Multiclass results with balanced datasets (testing)}\label{fig:13a_balanced}
\end{figure*}
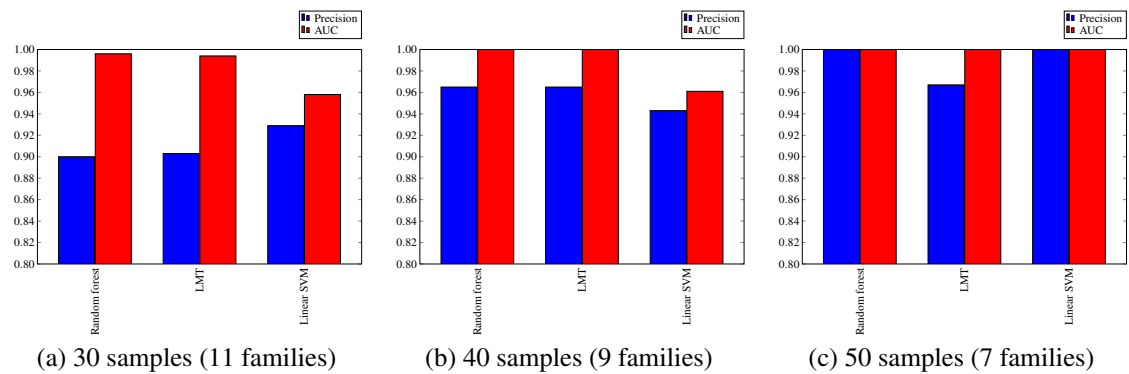

While there are~49 malware families in our dataset, 
from Figure~\ref{fig:10_fam_dist} we see that
a few large families dominate, while many of the ``families''
contain only~1 or~2 samples, and most are in the single digits.
To determine the effect of this imbalance on our results, 
we also performed multiclass experiments on balanced datasets. 
In Figures~\ref{fig:13a_balanced}~(a) through~(c) we give results for
balanced sets with~30, 40, and~50 samples per family, respectively.
For example, 11 of the~49 malware families in our dataset have at least~30
samples. From each of these~11 families, we randomly select~30 samples, then
perform multiclass experiments 
and we give the precision and AUC results in Figure~\ref{fig:13a_balanced}~(a).

Overall the results in Figure~\ref{fig:13a_balanced}
clearly demonstrate that our strong multiclass
results are not an artifact of the imbalance in our dataset.
In fact, with a sufficient number of samples per family,
we obtain better results on the balanced dataset than on the
highly imbalanced dataset. Of course, the number of families
is much smaller in the balanced case, but we do have a significant
number of families in all of these experiments.

Since a random forest performs best in the malware classification
problem, we give the confusion matrices for this case
in Figure~\ref{fig:15_train_conf_RF}. 
This confusion matrix provides a more fine-grained
view of the results reported above.
We observe that the
misclassification are sporadic, in the sense that
all of cases where a significant percentage of samples are
misclassified occur in small families, with the number
of samples being in the single digits.

\section{\uppercase{Conclusion}}\label{sect:con}

\noindent The work in~\cite{NC4} 
considered both permissions (i.e., a static feature) and system calls 
(i.e., a dynamic feature),
and the interplay between the two. The authors concluded that a slight reduction 
in the number of permissions has a significant effect, 
and suggested that malware writers may be able to evade detection by relatively 
minor modifications to their code, such as reducing the number of permissions requested.
Here, we provided a more nuanced analysis to show that it is likely
to be significantly more difficult to evade permission-based detection than 
suggested in~\cite{NC4}. 
Specifically, in this paper we showed that a relatively small number of permissions
can serve as a strong feature vector, even for the more challenging multiclass problem.
These results indicate that 
eliminating the specific permissions that comprise the reduced feature
set is likely not an option for most malware. In addition, we showed
that taking the opposite tack, that is, adding unnecessary permissions
that are common among bening apps, is also of limited value.  
We conclude that features based on permissions are likely to remain 
a viable option for detecting Android malware.

Our experimental results also show that malware detection on an Android device
is practical, since the necessary features can be extracted and scored efficiently.
For example, using an ANN on a reduced feature set, we can
obtain an AUC of~0.9920 for the binary classification problem.
And even in the case of highly skewed data---as would typically be expected
in a realistic scenario---an ANN can attain a testing accuracy 
in excess of~96\%.


The malware classification problem is inherently more challenging
than the malware detection problem. But even in this
difficult case, we obtained a testing accuracy of almost~95\%, based
on a random forest. It is worth noting that a random forest 
also performs well for binary classification, with about~97\%\  testing accuracy.
A random forest requires significantly less computing power to train, as
compared to an ANN, and this might be a factor in some implementations,
although training is often considered one-time work.

For future work, it would be interesting to further explore deep learning
for Android malware detection, based on permissions. For ANNs,
there are many parameters that can be tested, and it is possible that 
the ANN results presented in this paper can be significantly improved upon. 
As another avenue for future work, recent research has shown 
promising malware detection results by applying image analysis
techniques to binary executable files; 
see, for example~\cite{HuangTS18,YajamanamSTS18}.
As far as the authors are aware, such analysis has not been applied  
to the mobile malware detection or classification problems.

\bibliographystyle{apalike}
{\small
\bibliography{other.bib,Stamp-Mark.bib}}


\begin{figure*}[!htb]
	\input figures/fig_conf_train_per.tex
	\caption{Random forest confusion matrix as percentages (training)}\label{fig:15_train_conf_RF}
\end{figure*}







\end{document}